\documentclass[aps,preprint,amsmath,amssymb,showpacs]{revtex4}

\usepackage{graphicx}
\begin{document}

\title{Rare decay $Z \to \nu\bar{\nu}\gamma\gamma$ via tensor unparticle mediation}

\author{\.{I}nan\c{c} \c{S}ahin}
\email[]{isahin@science.ankara.edu.tr} \affiliation{Department of
Physics, Faculty of Sciences, Ankara University, 06100 Tandogan,
Ankara, Turkey}

\begin{abstract}
The decay width of the rare decay $Z \to \nu\bar{\nu}\gamma\gamma$
is strictly constrained from the LEP data. Tensor unparticles
provide a tree-level contribution to this rare decay. We have
calculated the tensor unparticle contribution to the rare decay $Z
\to \nu\bar{\nu}\gamma\gamma$. The current experimental limit have
been used to constrain unparticle couplings $\nu\bar{\nu}Z\cal
{U}^{\mu\nu}$ and $\gamma\gamma \cal {U}^{\mu\nu}$.
\end{abstract}

\pacs{14.80.-j, 12.90.+b, 13.38.Dg}

\maketitle

\section{Introduction}
Scale invariance plays a crucial role in theoretical physics. A
possible scale invariant hidden sector that may interact weakly with
the Standard Model (SM) fields is being discussed intensively in the
literature. Based on a scale invariant theory by Banks-Zaks (BZ)
\cite{BZ}, Georgi proposed a new scenario \cite{Georgi, Georgi2} in
which SM fields and a scale invariant sector described by (BZ)
fields interact via the exchange of particles with a large mass
scale $M_{U}$. Below this large mass scale interactions between SM
fields and BZ fields are described by non-renormalizable couplings
suppressed by powers of $M_{U}$ \cite{Georgi, Cheung}:

\begin{eqnarray}
\frac{1}{M_{U}^{d_{SM}+d_{BZ}-4}}O_{SM}O_{BZ}
\end{eqnarray}
The renormalization effects in the scale invariant BZ sector then
produce dimensional transmutation at an energy scale $\Lambda_{U}$
\cite{Weinberg}. In the effective theory below the scale
$\Lambda_{U}$, the BZ operators are embedded as unparticle
operators. The operator (1) match onto the following form,

\begin{eqnarray}
C_{O_{U}}\frac{\Lambda_{U}^{d_{BZ}-d_{U}}}{M_{U}^{d_{SM}+d_{BZ}-4}}O_{SM}O_{U}
\end{eqnarray}
here, $d_{U}$ is the scale dimension of the unparticle operator
$O_{U}$ and the constant $C_{O_{U}}$ is a coefficient function.

Phenomenological \cite{Luo}, astrophysical and cosmological
\cite{Davoudiasl} implications of unparticles have been intensively
studied in the literature. In the some of these phenomenological
researches several unparticle production processes have been
considered. A possible evidence for this scale invariant sector
might be a missing energy signature. It can be tested experimentally
by examining missing energy distributions. Another evidence for
unparticles can be explored by studying its virtual effects.

In this work we will present a detailed calculation of tensor
unparticle contribution to the rare decay $Z \to
\nu\bar{\nu}\gamma\gamma$. Experimental results from LEP data
\cite{booklet} strictly constrains the decay width at 95\%
confidence level:

\begin{eqnarray}
BR(Z \to \nu\bar{\nu}\gamma\gamma)\leq 3.1\times 10^{-6}
\end{eqnarray}
This experimental limit will be used to constrain unparticle
couplings $\nu\bar{\nu}Z\cal {U}^{\mu\nu}$ and $\gamma\gamma \cal
{U}^{\mu\nu}$.

\section{The rare decay $Z \to \nu\bar{\nu}\gamma\gamma$}

\begin{figure}
\includegraphics{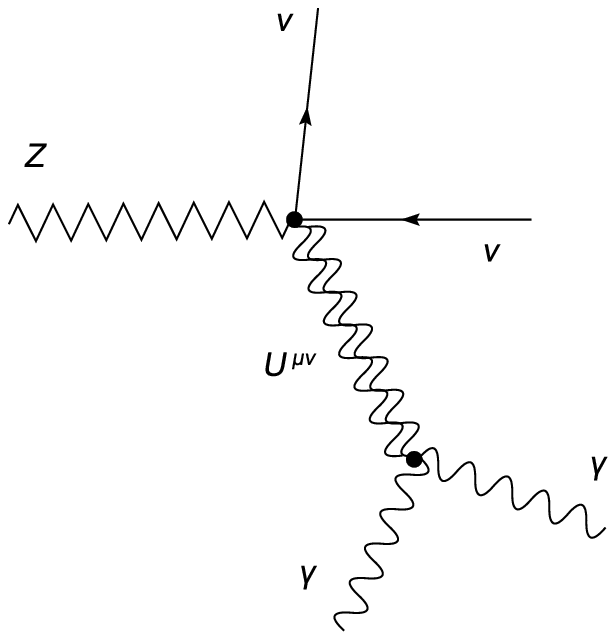}
\caption{Tensor unparticle contribution to the rare decay $Z \to
\nu\bar{\nu}\gamma\gamma$. \label{fig1}}
\end{figure}

$Z \to \nu\bar{\nu}\gamma\gamma$ decay may occur via tensor
unparticle exchange Fig.\ref{fig1}. We consider the following
effective interaction terms:

\begin{eqnarray}
-\frac{1}{4}\frac{\lambda}{\Lambda_{U}^{d_{U}}}\bar{\psi}i(\gamma_{\mu}D_{\nu}
+\gamma_{\nu}D_{\mu})\psi \cal{O_{U}^{\mu\nu}}\\
\frac{\kappa}{{\Lambda_{U}^{d_{U}}}}G_{\mu\alpha}
G_{\nu}^{\alpha}\cal{O_{U}^{\mu\nu}}
\end{eqnarray}
where
$D_\mu=\partial_\mu+ig\frac{\tau^{a}}{2}W_{\mu}^{a}+ig'\frac{Y}{2}B_{\mu}$
is the covariant derivative, $G^{\alpha\beta}$ denotes the gauge
field strength, $\psi$ is the Standard Model fermion doublet or
singlet, $\lambda$ and $\kappa$ are dimensionless effective
couplings. Feynman rules for these operators have been given in
ref.\cite{Cheung}. The vertex functions for
$\nu\bar{\nu}Z\cal{U}^{\mu\nu}$ and $\gamma(p_1)\gamma(p_2)
\cal{U}^{\mu\nu}$ generated from operators (4) and (5) are given by,

\begin{eqnarray}
&&\Gamma_{(\nu\bar{\nu}Z\cal{U})}^{\mu\nu\alpha}=\frac{ig}{4\cos\theta_W}\frac{\lambda}
{\Lambda_{U}^{d_{U}}}\left[\gamma^{\mu}g^{\nu\alpha}+\gamma^{\nu}g^{\mu\alpha}\right]\\
&&\Gamma_{(\gamma\gamma\cal{U})}^{\mu\nu\rho\sigma}=
\frac{i\kappa}{\Lambda_{U}^{d_{U}}}\left[K^{\mu\nu\rho\sigma}-K^{\mu\nu\sigma\rho}
\right]\\ \nonumber
&&K^{\mu\nu\rho\sigma}=-g^{\mu\nu}p_{1}^{\rho}p_{2}^{\sigma}-(p_{1}\cdot
p_{2})g^{\rho\mu}g^{\sigma\nu}+p_{1}^{\nu}p_{2}^{\rho}g^{\sigma\mu}+
p_{2}^{\mu}p_{1}^{\rho}g^{\sigma\nu}
\end{eqnarray}
respectively.

Spin-2 unparticle propagator is defined by \cite{Cheung}:

\begin{eqnarray}
\Delta(P^{2})_{\mu\nu,\rho\sigma}=i\frac{A_{d_{U}}}{2sin(d_{U}\pi)}(-P^{2})^{d_{U}-2}
T_{\mu\nu,\rho\sigma}(P)
\end{eqnarray}
where,

\begin{eqnarray}
A_{d_{U}}=\frac{16\pi^{\frac{5}{2}}}{(2\pi)^{2d_{U}}}\frac{\Gamma(d_{U}+\frac{1}{2})}
{\Gamma(d_{U}-1)\Gamma(2d_{U})}
\end{eqnarray}
\begin{eqnarray}
\pi_{\mu\nu}(P)=-g_{\mu\nu}+\frac{P_{\mu}P_{\nu}}{P^{2}}
\end{eqnarray}
\begin{eqnarray}
T_{\mu\nu,\rho\sigma}(P)=\frac{1}{2}\left[\pi_{\mu\rho}(P)\pi_{\nu\sigma}(P)+
\pi_{\mu\sigma}(P)\pi_{\nu\rho}(P)-\frac{2}{3}\pi_{\mu\nu}(P)\pi_{\rho\sigma}(P)\right]
\end{eqnarray}

The decay width can then be written as

\begin{eqnarray}
\Gamma(Z \to
\nu\bar{\nu}\gamma\gamma)=\frac{S}{2^{5}m_Z}\frac{1}{(2\pi)^{8}}\int
|\bar{\cal{M}}|^{2}\delta^{4}(Q-p_{1}-p_{2}-k_{1}-k_{2})
\frac{d^{3}p_{1}}{p_{1}^{0}}\,\frac{d^{3}p_{2}}{p_{2}^{0}}\,
\frac{d^{3}k_{1}}{k_{1}^{0}}\,\frac{d^{3}k_{2}}{k_{2}^{0}}
\end{eqnarray}
where $S=1/2!$ is the statistical factor for identical photons and
momentum 4-vectors of the participating particles are denoted as
$Z(Q) \to \nu(p_{1})\bar{\nu}(p_{2})\gamma(k_{1})\gamma(k_{2})$. By
applying Feynman rules we can obtain polarization summed squared
amplitude. It is given by

\begin{eqnarray}
|\bar{\cal{M}}|^{2}=\frac{g^{2}A_{d_{U}}^{2}}{m_{Z}^{2}\cos^{2}\theta_{W}\sin^{2}(d_{U}\pi)}
\left(\frac{\lambda}{\Lambda_{U}^{d_{U}}}\right)^{2}
\left(\frac{\kappa}{\Lambda_{U}^{d_{U}}}\right)^{2}
|(k_{1}+k_{2})^{2}|^{2d_{U}-4}\Theta_{\mu\nu}p_{1}^{\mu}p_{2}^{\nu}
\end{eqnarray}
where we have introduced the definition

\begin{eqnarray}
\Theta_{\mu\nu}=\left[(k_{1}\cdot k_{2})m_{Z}^{2}+2(k_{1}\cdot
Q)(k_{2}\cdot Q)\right]
\left(k_{1\nu}k_{2\mu}+k_{1\mu}k_{2\nu}\right)
\end{eqnarray}
In amplitude (13) we have assumed the lepton universality and a
factor of 3 has been taken into account for all of the known
neutrino species. Integration over $p_{1}$ and $p_{2}$ can be
provided with the aid of the following identity \cite{Singh}:

\begin{eqnarray}
I^{\mu\nu}=\int\frac{d^{3}p_{1}}{p_{1}^{0}}\frac{d^{3}p_{2}}{p_{2}^{0}}
\delta^{4}(V-p_{1}-p_{2})p_{1}^{\mu}p_{2}^{\nu}
=\frac{\pi}{6}\left(V^{2}g^{\mu\nu}+2V^{\mu}V^{\nu}\right)
\end{eqnarray}
Here $V=Q-k_{1}-k_{2}$. Then the decay width is reduced to the
following:

\begin{eqnarray}
\Gamma(Z \to
\nu\bar{\nu}\gamma\gamma)=\frac{1}{3\times2^{15}\pi^{7}m_{Z}^{3}}
\frac{g^{2}A_{d_{U}}^{2}}{\cos^{2}\theta_{W}\sin^{2}(d_{U}\pi)}
\left(\frac{\lambda}{\Lambda_{U}^{d_{U}}}\right)^{2}
\left(\frac{\kappa}{\Lambda_{U}^{d_{U}}}\right)^{2}\\
\nonumber\times
\int\frac{d^{3}k_{1}}{k_{1}^{0}}\frac{d^{3}k_{2}}{k_{2}^{0}}|(k_{1}+k_{2})^{2}|^{2d_{U}-4}
\,\Theta_{\mu\nu}\left(V^{2}g^{\mu\nu}+2V^{\mu}V^{\nu}\right)
\end{eqnarray}
There still remains the integrations over $k_{1}$ and $k_{2}$. These
integrations can be carried out numerically. We will work in the
center of mass system of the Z boson. We choose the following
integration variables:

\begin{eqnarray}
\xi=\frac{2Q\cdot k_{1}}{m_{Z}^{2}}=\frac
{2k_{1}^{0}}{m_{Z}},\,\,\,\,\,\,\, \eta=\frac{2Q\cdot
k_{2}}{m_{Z}^{2}}=\frac{2k_{2}^{0}}{m_{Z}},\,\,\,\,\,\,\,
\omega=\frac{(1-\cos\theta)}{2}
\end{eqnarray}
where $\theta$ is the angle between two outgoing photons. Then the
decay width can be written as

\begin{eqnarray}
\Gamma(Z \to \nu\bar{\nu}\gamma\gamma)=
\left(\frac{\lambda}{\Lambda_{U}^{d_{U}}}\right)^{2}
\left(\frac{\kappa}{\Lambda_{U}^{d_{U}}}\right)^{2}F(d_{U})
\end{eqnarray}

where,

\begin{eqnarray}
&&F(d_{U})=\frac{m_{Z}^{4d_{U}+1}}{3\times2^{16}\pi^{5}}
\frac{g^{2}A_{d_{U}}^{2}}{\cos^{2}\theta_{W}\sin^{2}(d_{U}\pi)}\\
\nonumber
\times&&\int_{\Omega}\xi^{3}\eta^{3}|\xi\eta\omega|^{2d_{U}-4}(1+\omega)\left[\omega(1+\xi\eta\omega-\xi-\eta)
+(1-\eta\omega)(1-\xi\omega)\right]d\xi d\eta d\omega
\end{eqnarray}

The integration region $\Omega$ is given by \cite{Singh,Perez}:

\begin{eqnarray}
0\leq\omega\leq1\,\,\,\,\,\,\,\,\,\,\,\,\,\,\,\,\,\,
\,\,\,\,\,\,\text{when}\,\,\,\,\,\,\,\,\,\,\,\,\,\,\,\,\,\,\,\,\,\,\,\,
\,\,\,\,\,\,0\leq\xi\leq1-\eta
\end{eqnarray}

\begin{eqnarray}
\frac{\xi+\eta-1}{\xi\eta}\leq\omega\leq1\,\,\,\,\,\,\,\,\,\,\,\,\,\,\,\,\,\,
\,\,\,\,\,\,\text{when}\,\,\,\,\,\,\,\,\,\,\,\,\,\,\,\,\,\,\,\,\,\,\,\,
\,\,\,\,\,\,1-\eta\leq\ \xi\leq1
\end{eqnarray}
together with $0\leq\eta\leq1$. Numerical integrations have been
performed by a Monte Carlo routine. During numerical integrations we
impose a cut $|cos\theta|<0.98$.

From current experimental limit (3) on the rare decay $Z \to
\nu\bar{\nu}\gamma\gamma$ we set the following bound:

\begin{eqnarray}
\left(\frac{\lambda}{\Lambda_{U}^{d_{U}}}\right)^{2}
\left(\frac{\kappa}{\Lambda_{U}^{d_{U}}}\right)^{2}F(d_{U})\leq7.73512\times10^{-6}
\end{eqnarray}
This expression gives us the allowed region in the
$\frac{\lambda^{2}}{\Lambda_{U}^{2d_{U}}}$ versus
$\frac{\kappa^{2}}{\Lambda_{U}^{2d_{U}}}$ plane. In Fig.\ref{fig2},
we plot the boundary lines of scaled unparticle couplings
$\frac{\lambda^{2}}{\Lambda_{U}^{2d_{U}}}$ and
$\frac{\kappa^{2}}{\Lambda_{U}^{2d_{U}}}$ for various values of the
scale dimension $d_{U}$. Allowed regions are defined by the area
restricted by these boundary lines.

In principle, unparticle couplings $\kappa$ and $\lambda$ can be
different. If we assume $\kappa=\lambda$, we can obtain a unique
bound on $\lambda$ for each value of the scale dimension. The bounds
on scaled unparticle couplings
$|\frac{\lambda}{\Lambda_{U}^{d_{U}}}|$
 are given on Table \ref{tab1}.

\section{conclusion}

\begin{figure}
\includegraphics{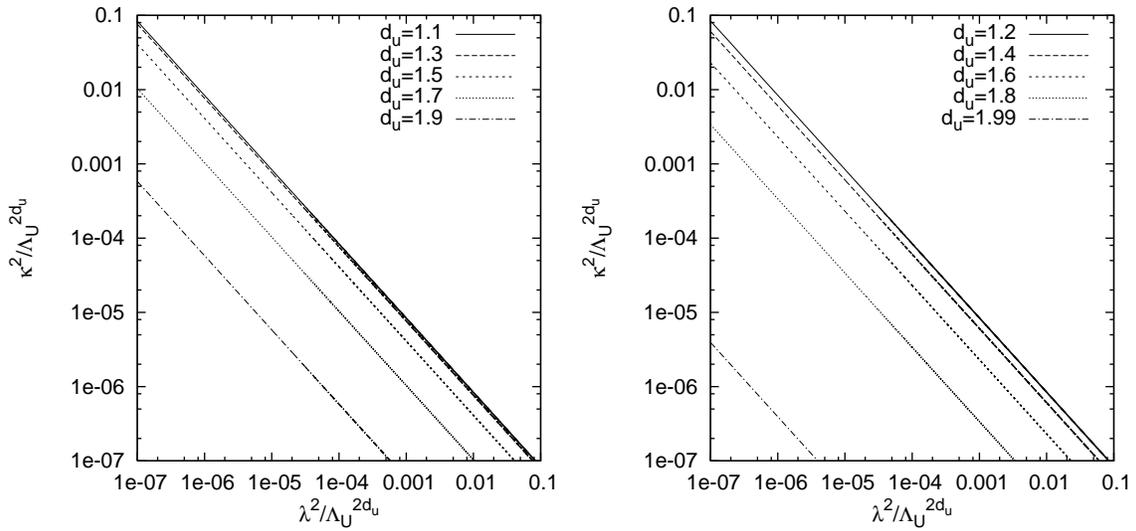}
\caption{Boundary lines of scaled unparticle couplings
$\frac{\lambda^{2}}{\Lambda_{U}^{2d_{U}}}$ and
$\frac{\kappa^{2}}{\Lambda_{U}^{2d_{U}}}$. Legends are for various
values of the scale dimension $d_{U}$. Allowed regions are defined
by the area restricted by the boundary lines. \label{fig2}}
\end{figure}

One can see from Fig.\ref{fig2} that allowed area decreases in size
as $d_{U}$ increases. This feature is reflected in the Table
\ref{tab1} also. For instance, bound on
$|\frac{\lambda}{\Lambda_{U}^{d_{U}}}|$  decreases by a factor of
11.9 as $d_{U}$ increases from 1.01 to 1.99 (Table \ref{tab1}). This
behavior is clear from factor $\frac{1}{\sin^{2}(d_{U}\pi)}$ in the
squared amplitude (13).

Tensor unparticle interactions (4) and (5) also contribute to the
rare decay $Z \to \ell \bar{\ell}\gamma\gamma$. Experimental limit
on this rare decay is at the same order $O(10^{-6})$ as $Z \to
\nu\bar{\nu}\gamma\gamma$ decay. Therefore this experimental limit
can also be used to constrain tensor unparticle couplings. The
difference is that the decay $Z \to \ell \bar{\ell}\gamma\gamma$
receives Standard Model tree-level contributions but $Z \to
\nu\bar{\nu}\gamma\gamma$ decay does not.


\pagebreak

\begin{table}
\caption{Bounds on tensor unparticle couplings. \label{tab1}}
\begin{tabular}{cc}
Scale dimension $d_{U}$ & \,\,\,\,\,\,\,\,\,\,\,\,
\,\,\,\,Bounds on $|\frac{\lambda}{\Lambda_{U}^{d_{U}}}|$\\
\hline
$d_{U}=1.01$&\,\,\,\,\,\,\,\,\,\,\,\,\,\,\,\,
 $|\frac{\lambda}{\Lambda_{U}^{d_{U}}}|\leq0.00938$ \\
$d_{U}=1.1$&\,\,\,\,\,\,\,\,\,\,\,\,\,\,\,\,
 $|\frac{\lambda}{\Lambda_{U}^{d_{U}}}|\leq0.00956$ \\
$d_{U}=1.2$&\,\,\,\,\,\,\,\,\,\,\,\,\,\,\,\,
 $|\frac{\lambda}{\Lambda_{U}^{d_{U}}}|\leq0.00958$ \\
$d_{U}=1.3$&\,\,\,\,\,\,\,\,\,\,\,\,\,\,\,\,
 $|\frac{\lambda}{\Lambda_{U}^{d_{U}}}|\leq0.00935$ \\
$d_{U}=1.4$&\,\,\,\,\,\,\,\,\,\,\,\,\,\,\,\,
 $|\frac{\lambda}{\Lambda_{U}^{d_{U}}}|\leq0.00883$ \\
$d_{U}=1.5$&\,\,\,\,\,\,\,\,\,\,\,\,\,\,\,\,
 $|\frac{\lambda}{\Lambda_{U}^{d_{U}}}|\leq0.00800$ \\
$d_{U}=1.6$&\,\,\,\,\,\,\,\,\,\,\,\,\,\,\,\,
 $|\frac{\lambda}{\Lambda_{U}^{d_{U}}}|\leq0.00693$ \\
$d_{U}=1.7$&\,\,\,\,\,\,\,\,\,\,\,\,\,\,\,\,
 $|\frac{\lambda}{\Lambda_{U}^{d_{U}}}|\leq0.00567$ \\
$d_{U}=1.8$&\,\,\,\,\,\,\,\,\,\,\,\,\,\,\,\,
 $|\frac{\lambda}{\Lambda_{U}^{d_{U}}}|\leq0.00429$ \\
$d_{U}=1.9$&\,\,\,\,\,\,\,\,\,\,\,\,\,\,\,\,
 $|\frac{\lambda}{\Lambda_{U}^{d_{U}}}|\leq0.00276$ \\
$d_{U}=1.95$&\,\,\,\,\,\,\,\,\,\,\,\,\,\,\,\,
 $|\frac{\lambda}{\Lambda_{U}^{d_{U}}}|\leq0.00185$ \\
$d_{U}=1.99$&\,\,\,\,\,\,\,\,\,\,\,\,\,\,\,\,
 $|\frac{\lambda}{\Lambda_{U}^{d_{U}}}|\leq0.00079$ \\
\end{tabular}
\end{table}

\end{document}